# Modified Sagnac experiment for measuring travel-time difference between counter-propagating light beams in a uniformly moving fiber

Ruyong Wang [a], Yi Zheng [b,*], Aiping Yao [b], Dean Langley [c]

[a]*St. Cloud State University, St. Cloud, MN 56301, USA*
[b]*Department of Electrical and Computer Engineering, St. Cloud State University, St. Cloud, MN 56301, USA*
[c]*Physics Department, St. John's University, Collegeville, MN 56321, USA*

**Abstract**

A fiber optic conveyor has been developed for investigating the travel-time difference between two counter-propagating light beams in uniformly moving fiber. Our finding is that there is a travel-time difference $\Delta t = 2v\Delta l/c^2$ in a fiber segment of length $\Delta l$ moving with the source and detector at a speed $v$, whether the segment is moving uniformly or circularly.

## 1. Introduction

The Sagnac effect [1] shows that two light beams, sent clockwise and counterclockwise around a closed path on a rotating disk, take different time intervals to travel the path. The time difference between them is given generally by $\Delta t = 4A\Omega/c^2$, where $A$ is the area enclosed by the path, $\Omega$ is the angular velocity of the rotation [2]. For a circular path of radius $R$, the difference can also be represented as $\Delta t = 2vl/c^2$, where $v = \Omega R$ is the speed of the circular motion and $l = 2\pi R$ is the circumference of the circle. The Sagnac effect is a first order effect in $v/c$ and it has been found in many systems ranging in size from the around-the-world Sagnac experiment [3] to small fiber optic gyroscopes (FOGs) [4]. In a FOG, when a single mode fiber is wound to a coil with $N$ turns, the Sagnac effect is enhanced to $\Delta t = 4A\Omega N/c^2$ or $\Delta t = 2vL/c^2$, where $L$ is the fiber length. The travel-time difference in the FOG can also be expressed by the phase shift, $\Delta\phi = 2\pi\Delta tc/\lambda$, where $\lambda$ is the free space wavelength of light.

The existence and importance of the Sagnac effect have been well acknowledged, although a dozen interpretations can be found [5]. A rotating frame of reference is usually used in explanations, but Sagnac himself interpreted the effect without the use of relativity [1], while general relativity is used by others [6]. We designed a modified Sagnac experiment to examine whether the travel-time difference only appears in rotational motion, or if it also appears in uniform motion. Our experiment is important because it shows that the time-difference effect also occurs for uniform motion.

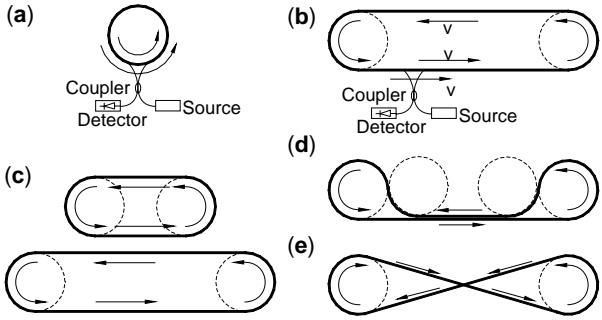

Fig. 1. (a) A FOG including a source and detector rotating with the fiber, is transformed into (b) a FOC in a regular configuration. (c) Regular configuration with different lengths for straight-fiber segments, (d) zero-area configuration and (e) figure-8 configuration.

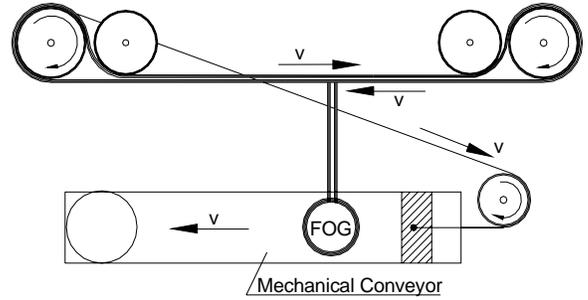

Fig. 2. Experimental setup. The mechanical conveyor is used to move the fiber optic loop and carry the light source and detector in the FOG at the same speed.

## 2. Method

Conceptually, a FOG, shown in Fig. 1(a), could be divided into two semicircular sections with extended fiber connecting the end sections as shown in Fig. 1(b). The fiber moves when the wheels at the two ends rotate. We call this new device a Fiber Optic Conveyor (FOC). The travel-time difference due to uniform motion was investigated by using different lengths for the straight-fiber segments with the same semicircular end sections. As a matter of fact, in a FOC, the travel-time difference between two beams through the loop is a summation of all the travel-time differences between two beams in each segment of the loop. Therefore, the comparison of the two travel-time differences of the two FOCs, that have the same semicircular end sections but different lengths for the straight-fiber segments as shown in Fig. 1(c), will reveal whether or not the travel-time difference exists in uniformly moving segments of the fiber. As in the case of the FOG, the FOC carries multiple turns of fiber to increase the travel-time difference.

If the travel-time difference appears in uniform motion, it should be independent of the enclosed area that is a determining factor in the Sagnac effect caused by rotational motion. Therefore, we designed two additional conveyor configurations: zero-area (Fig. 1d) and figure-8 (Fig. 1e). In the zero-area configuration, the enclosed area of the main part of the conveyor is almost zero except the two ends of the conveyor. The two counter-moving linear segments of the fiber are like two shoulder-brushing trains on parallel tracks. In the figure-8 configuration, from the point of view of rotation, the two enclosed areas have opposite directions so the effective enclosed area of the loop is approximately zero. This means if we conduct the Sagnac experiment by putting the figure-8 configuration on a rotating disk, the phase shifts of the two counter-wound portions cancel each other.

Our experiments differ from the Fizeau type experiment [7] in which the medium, water or glass, is moving, but the light source and detector are stationary. In our FOC experiment, as in the Sagnac type experiments, the light source and the detector are co-moving with the medium.

## 3. Experiment

The experiment was conducted using a FOC that was constructed by a FOG [8] which was modified by adding an extra 50 m of single mode fiber to the original fiber loop (Fig. 2). The extra fiber was wrapped onto a polyester ribbon that went around two wheels with diameters of 30 cm to form our FOC. This fiber loop was moved by one wheel dragged by a 1.5-m long mechanical conveyor. For the zero-area configuration, additional wheels pushed inward on the loop so the middle portion of the FOC enclosed very little area. For the figure-8 configuration the ribbon was twisted between the two ends. The FOG travelled with the mechanical conveyor and its uniform motion did not cause any phase shift because the FOG is only sensitive to the rotational movement. The light source for the FOG was a 1310-nm superluminescent LED. The FOG with the extended fiber was calibrated and its output rate is 1162.6 mV per radian of detected phase shift. The phase shift of the FOG is linearly

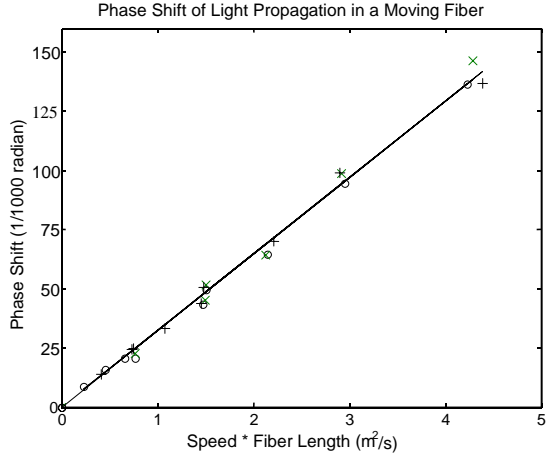

Fig. 3. FOC phase shifts versus products of fiber speeds and lengths. Symbol + is for the zero-area configuration, × for the figure-8 configuration, and o for the regular configuration. The straight line was found by a least square linear regression of all measurements.

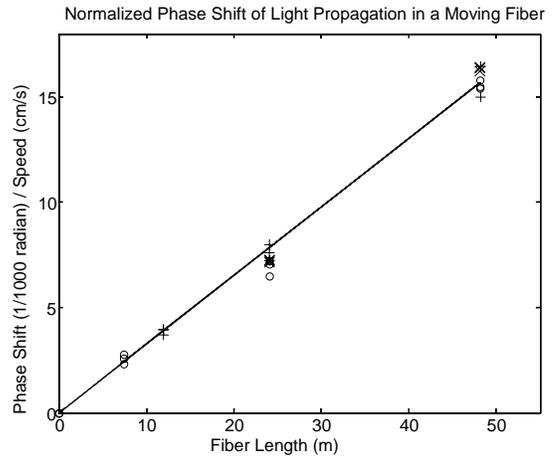

Fig. 4. Speed-normalized phase shifts versus fiber lengths. The symbols used are the same as in Fig. 3. The straight line was found by a least square linear regression of all data.

proportional to $vL$ at a constant rate of 0.03200 radians per $m^2$/s.

The experiment was repeated with 24 different arrangements of conveyor speeds, fiber lengths, and the three different FOC configurations shown in Fig.1. The conveyor speeds were between 3 and 9 cm/s. The loops had perimeters of 2.5, 4.0, 8.0, and 16.0 m; in each case there were three turns of the fiber wound on the loop. The remainder of the 50 m of fiber was wound on a coil of 9 cm in diameter which moved uniformly with the FOG so as not to produce any phase shift. When the conveyor ran at various stable speeds, 500 to 1,500 samples were recorded for each measurement. Measurements were repeated about eight times for each of the 24 arrangements of the FOC. Means and standard deviations of phase shifts were calculated for each measurement, and these were averaged for each of the 24 arrangements. Those means of the phase shifts are shown in Fig. 3 and are linearly proportional to the speed of moving fiber in the laboratory and fiber length. Least-square linear regression values of the FOC output rates are 0.03200 ± 0.00528, 0.03300 ± 0.00274, and 0.03171 ± 0.00282 radians per $m^2$/s for zero-area, figure-8, and regular configurations, respectively, and 0.03229 ± 0.00365 radians per $m^2$/s for the data as a whole; these are all in agreement with the FOG output rate due to pure rotation: $\Delta\phi = 4\pi vL/c\lambda = 0.03200 vL$.

**4. Discussion**

As shown in Fig. 3, the phase shift or the travel-time difference between two counter-propagating light beams in the moving optic fiber was clearly observed in all different configurations of FOCs. The phase shift $\Delta\phi$, and therefore, the travel-time difference $\Delta t$ are proportional to both the total length and the speed of the moving fiber whether the motion is circular or uniform. Other tests using smaller end wheels for the FOC and fiber loops with additional curves also confirmed the same finding. To test the linearity of the phase shift to $v$ and $L$ independently, the phase shifts were normalized by the speeds and were found to be linearly proportional to the moving fiber length in Fig. 4.

It is interesting to notice that when the straight-fiber segment is made shorter and shorter, a FOC with the regular configuration reduces to a FOG, and a FOC with the figure-8 configuration reduces to two counter-wound, counter-rotating FOGs, of which the phase shifts augment each other. While the phase shift in the Sagnac effect is often expressed using the enclosed area as a factor, our results indicate that the length and speed of the moving fiber are the fundamental factors, rather than the enclosed area.

Finally, we found the phase shift due to uniform motion of a fiber segment was linearly proportional to its length, by subtracting the $\Delta\phi$ of the shortest loop from each $\Delta\phi$ of longer loops for a given speed. We conclude that a segment of uniformly moving fiber with a speed of $v$ and a length of $\Delta l$ contributes $\Delta\phi = 4\pi v\Delta l/c\lambda$ or $\Delta t = 2v\Delta l/c^2$, like a segment of circularly moving fiber does. This travel-time difference observed in our experiment is another first order effect in $v/c$, like the Sagnac effect.

The travel-time difference $\Delta t = 2v\Delta l/c^2$ in our experiment is independent of the refractive index $n$, as in the Sagnac type experiment. It implies that the same should occur in vacuum (refractive index $n = 1$), which could be tested by using a FOC with a hollow-core single-mode fiber in which light is guided in vacuum or in air [9]. The analysis of the implication of travel-time difference in a uniformly moving vacuum light-guide to the propagation of light in vacuum will be very interesting.

## 5. Conclusion

The travel-time difference of two counter-propagating light beams in moving fiber is proportional to both the total length and the speed of the fiber, regardless of whether the motion is circular or uniform. In a segment of uniformly moving fiber with a speed of $v$ and a length of $\Delta l$, the travel-time difference is $2v\Delta l/c^2$.

## Acknowledgements

We thank Robert Moeller of the Naval Research Laboratory for technical assistance and NRL for the loan of the FOG. We thank Xiaonan Shen for assistance.

* Corresponding author. *E-mail address*: zheng@stcloudstate.edu